\documentclass[12pt,preprint]{aastex}

\shorttitle{NEAR-INFRARED SPECTROSCOPY OF SDSS 1212}
\shortauthors{Farihi, Burleigh, \& Hoard}

\begin{document}

\title{A NEAR-INFRARED SPECTROSCOPIC STUDY OF
		THE ACCRETING MAGNETIC WHITE DWARF 
		SDSS J121209.31+013627.7 AND ITS SUBSTELLAR COMPANION}

\author{J. Farihi\altaffilmark{1,2},
	 	M. R. Burleigh\altaffilmark{3}, \& 
	 	D. W. Hoard\altaffilmark{4}}

\altaffiltext{1}{Gemini Observatory,
			Northern Operations,
			670 North A'ohoku Place,
			Hilo, HI 96720}
\altaffiltext{2}{Department of Physics \& Astronomy,
			University of California,
			430 Portola Plaza,
			Los Angeles, CA 90095; jfarihi@astro.ucla.edu}
\altaffiltext{3}{Department of Physics \& Astronomy,
			University of Leicester,
			Leicester LE1 7RH, UK; mbu@star.le.ac.uk,}
\altaffiltext{4}{Spitzer Science Center,
			California Institute of Technology,
			MS 220-6,
			Pasadena, CA 91125; hoard@ipac.caltech.edu}

\begin{abstract}

The nature of the excess near-infrared emission associated with the magnetic white dwarf 
commonly known as SDSS 1212 is investigated primarily through spectroscopy, and also via 
photometry.  The inferred low mass secondary in this system has been previously detected by 
the emission and variation of H$\alpha$, and the $1-2.5$ $\mu$m spectral data presented here 
are consistent with the presence of a late L or early T dwarf.  The excess flux seen beyond 1.5 
$\mu$m in the phase-averaged spectrum is adequately modeled with an L8 dwarf substellar 
companion and cyclotron emission in a 7 MG magnetic field.  This interesting system manifests 
several observational properties typical of polars, and is most likely an old interacting binary with 
a magnetic white dwarf and a substellar donor in an extended low state.

\end{abstract}

\keywords{binaries: close---infrared: stars---stars: 
	fundamental parameters---stars: individual 
	(SDSS J121209.31+013627.7)---stars: low-mass, 
	brown dwarfs---novae, cataclysmic variables---stars:
	evolution---stars: formation---white dwarfs}

\section{INTRODUCTION}

Cataclysmic variables are spectrally exotic binaries, displaying a diversity and complexity 
of features, behavior, and variability, effectively masking the spectral signatures of their coolest 
secondaries.  Despite unyielding efforts to unambiguously and directly detect likely substellar 
companions in either non-magnetic or magnetic cataclysmic variables, these altered degenerate
dwarfs remain largely elusive \citep*{lit03}.  The mass determination of the substellar donors in 
two non-magnetic accreting binaries reported by \citet*{lit06} and \citet*{lit07} remain rare among 
cataclysmic variables, although the determination of their substellarity was indirect.  Cataclysmic 
variables containing strongly magnetic white dwarfs (called polars because of the highly polarized 
nature of their light) offer a slight advantage by lacking the often overwhelming luminosity from 
an accretion disk.  On the other hand, their magnetic fields generate copious cyclotron radiation 
(in addition to flux from direct accretion), which is not present in non-magnetic cataclysmic variables.

Polars are also known for exhibiting long periods of quiescence in which many secondaries 
have been directly observed, although none of spectral type later than M \citep*{har05}.  The 
recently recognized class of low accretion rate polars may provide decent infrared hunting 
grounds for substellar companions, if present, as their cyclotron harmonics appear mainly in 
the optical due to $B\sim50$ MG fields.  It has been suggested that no mass transfer takes place 
in these systems, but rather the efficient capture of secondary wind, and that their separations are 
consistent with being essentially detached \citep*{sch05b}.  Yet, it is difficult to certify that these 
binaries are not simply in protracted low states such as seen in other polars \citep*{sch07,har04}.

Recent {\em Spitzer} IRAC observations of several polars indicate flat infrared spectral energy 
distributions which are difficult to reconcile with cool degenerate companions alone, and may 
indicate circumbinary dust as well as underlying cyclotron continuum emission \citep*{bri07,
how06}.  It will be interesting to see the results of similar mid-infrared observations with a 
larger sample of polars as well as observations at longer wavelengths where there should 
be little or no cyclotron emission even in $B\approx10$ MG fields.  The first published {\em 
Spitzer} IRS spectrum of a polar strongly supports the presence of circumbinary dust at EF 
Eridani \citep*{hoa07b}.

This paper represents a spectroscopic effort to directly detect the almost certain substellar 
companion to the cool magnetic white dwarf SDSS J121209.31+013627.7 (SDSS 1212).  Its 
binarity was reported with optical spectra supplemented with a single $J$-band photometric 
observation, limiting the spectral type of the secondary to L5 or later, and was initially thought to 
represent a detached system \citep*{sch05a}.  This possibility was still considered viable despite 
a second study which detected strong cyclotron emission -- a likely indicator of mass transfer -- in 
$K$-band time series photometry \citep*{deb06}.  Both \citet*{koe06} and \citet*{bur06a} detected 
optical variability and a light curve similar to that shown by polars in low states.  \citet*{bur06a}
further showed that the amplitude of the modulations was larger at blue and ultraviolet 
wavelengths than in the red, consistent with a hot accretion spot responsible for the cyclotron 
emission.  \citet*{bur06a} also detected x-ray emission from SDSS 1212, conclusively proving 
accretion is taking place.  This paper presents cross-dispersed $1-2.5$ $\mu$m spectroscopy
and spectral modeling, along with $HK$ photometry, in order to constrain the source(s) of the 
observed near-infrared excess emission.

\section{OBSERVATIONS}

\subsection{$HK$ Photometry}

Near-infrared images of SDSS 1212 were acquired on 2006 June 10 at the United Kingdom 
Infrared Telescope under photometric conditions ($\sec{z}<1.1$) with the Wide Field Camera 
(WFCAM; Casali et al., in preparation; \citealt*{hir06,hen03}).  A 5 point dither pattern was used 
with $2\times10$ s exposures at each position; this pattern was repeated twice at $H$ and thrice 
at $K$ for total exposure times of 200 and 300 s, respectively.  The data were reduced in the 
standard fashion.  Each stack was median combined to extract a sky frame which was also 
normalized to generate a flat field map.  Individual sky subtracted, flat fielded frames were 
registered and averaged for a final frame upon which to execute measurements.  The full 
widths at half maximum of the reduced stellar images were 2.0 and 1.9 WFCAM pixels ($0''.40$ 
pixel$^{-1}$) at $H$ and $K$ respectively.  Aperture photometry was performed with standard 
IRAF tasks using an $r=1''.20$ aperture and an $r=8-12''$ sky annulus.

Due to the very wide field of view of WFCAM ($819''\times819''$), 15 2MASS point sources 
in the same field as SDSS 1212 were simultaneously imaged to choose amongst as possible 
photometric calibrators.  Each 2MASS star chosen as a calibrator had all of the following 
properties: a full width at half maximum consistent with linear accumulation (i.e. unsaturated), 
high signal-to-noise ratios (S/N $>$ 100) in the reduced WFCAM images, and reliable 2MASS 
photometry at both wavelengths ($\sigma_{H}$ $<$ 3.4\%, $\sigma_{K_s}$ $<$ 6.3\%; 
\citealt*{skr06}).  The selected 12 calibrators are listed in Table \ref{tbl1}, along with their 
2MASS data and their derived instrumental zeropoints.

Additional photometry was acquired on 3 July 2006 at the Gemini Telescope North under 
photometric conditions with the Near-Infrared Imager (NIRI; \citealt*{hod03}).  Both SDSS 1212 
and the UKIRT standard star FS 132 \citep*{haw01} were observed in $HK$ at nearly identical 
average airmasses of $\sec{z}\approx1.95$, yielding extinction corrections less than 0.003 mag.
A 5 point dither pattern was used for both science target and calibrator, with an extra frame taken 
for SDSS 1212.  The calibrator was observed $10\times1$ s at each position while the science 
target was observed $3\times12$ s at each position, yielding total exposure times of 50 s for FS 
132 and 216 s for SDSS 1212 at $HK$.  The data were reduced in a manner similar to the WFCAM 
data, with separately acquired and processed flat field images. The full widths at half maximum of 
the reduced stellar images were between 4.5 and 4.9 NIRI pixels ($0''.12$ pixel$^{-1}$ at f/6).  
Aperture photometry was performed similarly using an $r=0''.47$ aperture and an $r=2''.3-3''.5$ 
sky annulus.

The photometry results are listed in Table \ref{tbl2} and plotted in Figure \ref{fig1}.  The third 
column in the table gives the uncertainty in the measured aperture photometry (due to intrinsic 
S/N limitations).  The fourth column in the table gives the calibration uncertainty: for the WFCAM 
data, this is the standard deviation in the derived zeropoints; for the NIRI data, an intrinsic 3\% 
error was assumed.

\subsection{$1-2.5$ $\mu$m Spectroscopy}

Spectroscopy was performed on 2006 June 19 and 23 at the Gemini Telescope South 
with the Near-Infrared Spectrograph (GNIRS; \citealt*{eli98}).  Observations were executed in 
cross-dispersed mode with the 31.7 l/mm grating and the $0''.3$ wide slit, yielding $R\approx
1700$ while effectively covering the entire near-infrared range without inter-order contamination.  
Data were taken at two nod positions along the $6''$ long slit, with 300 s exposures at each nod 
position, repeated 6 times for a total on source integration time of 1 hr on each of two nights.  The 
telluric standard HIP 60030 (A7V, $V-K=0.5$) was observed in a similar manner with 6 s exposures 
at each nod position, repeated 4 times.

The raw data were processed with standard GNIRS tools within the Gemini IRAF package 
version 1.9.  These steps included flat field and pinhole map creation from calibration lamps, 
bad pixel correction, linearity correction, sky subtraction, image shifting and averaging, order 
cutting, spatial rectification, and wavelength calibration from argon lamps.  Spectral extraction 
of each order was performed with a GNIRS equivalent to the IRAF {\sf apall} task.

Some processing of the raw data was required prior to application of the standard package tools
listed above.  The first and second science frames on each night contained significant residual 
signal (between 3 and 5 times larger than the read level in the unexposed portion of the array) 
from the target acquisition mirror.  The raw acquisition images were scaled to match the 
contamination level of the affected frames and subtracted off, effectively solving this problem.  
Additionally, there is an occasionally occurring bias level problem in one quadrant of the detector 
which appeared in many of the science exposures.  This noise was well removed with the {\sf 
nvnoise} task within the GNIRS tools.

\begin{figure}
\plotone{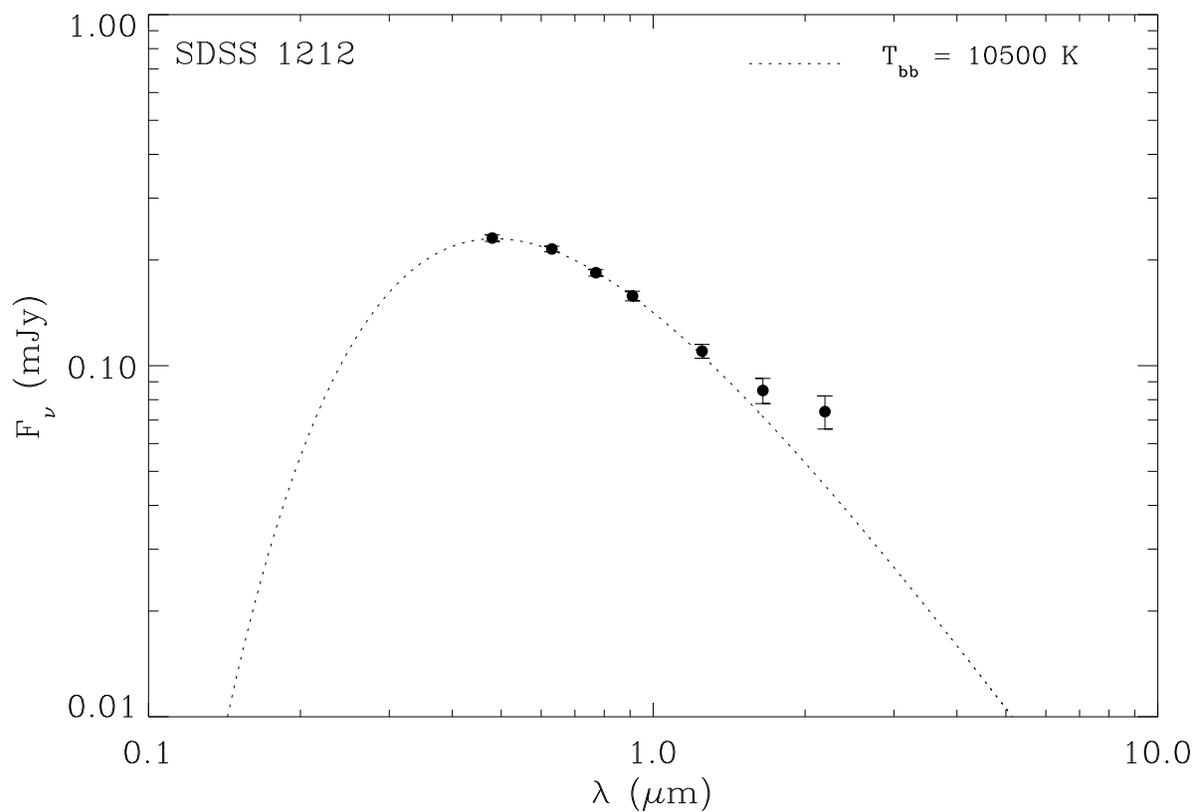}
\caption{Spectral energy distribution of SDSS 1212.  Optical $gri$ and near-infrared 
$zJ$ photometry is from \citet*{sch05a}.  The $H$-band value is the average of the NIRI 
and WFCAM measurements, while the $K$-band value is the flux measured during 
minimum (WFCAM), all from the present paper (Table \ref{tbl2}).  The $grizJ$ photometric 
fluxes are well reproduced by a 10,500 K blackbody.
\label{fig1}}
\end{figure}

Both the telluric standard and SDSS 1212 were processed in an identical fashion until the 
spectral extraction step.  Instead of the $8-10$ pixel diameter ($1''.20-1''.50$) aperture used for 
HIP 60030, a $4-5$ pixel diameter ($0''.60-0''.75$) aperture was used for SDSS 1212 to maximize 
the extracted spectroscopic S/N of this faint source.  Flux calibration and telluric correction were 
performed for each order with Spextool (version 3.3; \citealt*{cus04,vac03}), using the general 
{\sf xtellcor} package.  The flux levels of all four spectral orders presented here agree quite well 
at their boundaries and hence no inter-order adjustment was necessary.  However, the entire 
cross-dispersed spectrum, the average of both nights' data, was adjusted slightly by a single 
scaling factor to best match the $J$ and $H$-band photometry in Figure \ref{fig1}.  This 
unsmoothed spectrum is displayed in Figure \ref{fig2} together with the $JHK$ points from 
Figure \ref{fig1}.

\section{RESULTS}

When telescope time was requested for this study, it was not yet known that SDSS 1212 was 
variable in any broad wavelength bands.  Apart from the H$\alpha$ line strength variability 
reported by \citet*{sch05a}, SDSS 1212 is now known to be variable in $ugi$, $VR$ (and 
unfiltered white optical light), and $K$, but not in $JH$ \citep*{bur06a,koe06,deb06}.

Figure \ref{fig3} shows the corresponding phase of SDSS 1212 for all the photometric and 
spectroscopic observations.  The WFCAM photometry was taken during $K$-band minimum, 
by which it is meant that all the 2.2 $\mu$m flux should be near its minimum value -- whether 
this minimum flux comes from cyclotron continuum emission, the substellar companion or 
a combination of sources is uncertain at present and the focus of this study.  The NIRI 
photometry, on the other hand, was taken close to the $K$-band maximum which occurs 
near orbital phases $\phi=0.8$ and 0.2, but not at $\phi=0.0$ as for the optical and ultraviolet 
peak (see \S3.2).  

\subsection{WFCAM Photometry During Minimum}

It is somewhat surprising that the nominal WFCAM $H$ and $K$ magnitudes in Table
\ref{tbl2} are $0.12-0.15$ mag {\em fainter} than the minimum reported by \citet*{deb06}, 
although the datasets have overlapping error bars at each passband.  As can be seen in 
Figure \ref{fig3}, it has been verified that the WFCAM observations were taken during the 
photometric minimum by comparing the corresponding corrected MJD with the ephemeris 
of \citet*{bur06a}.

Five of the stars listed in Table \ref{tbl1} were used as calibrators by \citet*{deb06}.  
However, the 2MASS photometric errors on 3 of these objects, while mainly decent at $J$, 
become relatively high at $H$ and $K_s$ ($\sigma_J$ $<$ 11\%, $\sigma_H$ $<$ 17\%, 
$\sigma_{K_s}$ $<$ 25\%), and were discarded here for this reason.  In order to investigate 
whether this fact could account for the discrepancy between the reported $HK$ magnitudes 
-- and more importantly, to assess the likelihood that this difference is real -- the instrumental 
zeropoints for the 3 discarded calibrators were compared with those derived with the 12 higher 
S/N stars used here.  Those 3 stars alone yield an $H_0$ and $K_0$ which differ by 3 standard 
deviations ($-0.06$ mag at $H$, $-0.14$ mag at $K$) from the mean zeropoints used for the 
Table \ref{tbl2} WFCAM results.  Additionally, according to \citet*{deb06}, not all 5 of these 
calibrators were present in every frame of their WIRC and PANIC data.  The 3 stars with lowest 
S/N (those discarded here) are in fact closest on the sky to SDSS 1212, at $61-93''$, while the 
remaining 2 calibrators are $115-126''$ distant.  Hence the 3 lower quality 2MASS calibrators 
had a good chance of being present in most, if not all, the WIRC frames (a single chip is $201''
\times 201''$), while the 2 higher quality calibrator stars were unlikely to be imaged in the majority 
of that dataset.  Owing to the smaller field of view of PANIC ($128''\times128''$), only the 3 lowest 
S/N 2MASS calibrators were used \citep*{deb06}.

There are two additional reasons why the WFCAM magnitudes in Table \ref{tbl2} may be 
slightly discrepant from the previously published values.  First, neither study has transformed the 
photometry from the 2MASS system into the native filter system through which the observations 
were taken.  In the present case, any errors introduced by failing to transform the photometry are 
much smaller than the errors due to the intrinsic S/N (represented by $\sigma_{phot}$ in Table 
\ref{tbl2}).  Also, the fact that 12 calibrators were used here would almost certainly shrink any 
error of this type even further due to averaging \citep*{car01}.  Second, it is conceivable that 
SDSS 1212 exhibits further variability which has so far not been documented.

\subsection{NIRI Photometry Near Maximum}

The NIRI photometry is also listed in Table \ref{tbl2}.  It has been verified that these observations 
were taken near the minimum hot spot viewing angle.  Although this corresponds to a maximum 
in ultraviolet and optical light, it does not quite correspond to the cyclotron maximum as seen in
the $K$-band light curve measured by \citet*{deb06}, which occurs when the hot spot is on either 
limb -- immediately before and after the ultraviolet maximum ($\phi=0.0$ in this paper and 
\citealt*{bur06a}, $\phi=0.5$ in \citealt*{deb06}).

\subsection{GNIRS Phase-Averaged Spectroscopy}

The spectroscopy essentially covers all orbital phases of SDSS 1212 because the flux was 
sufficiently low (0.1 mJy at 1.6 $\mu$m) as to preclude extracting phase-resolved data -- the 
observation itself represents a benchmark near the limit of what is possible with the most 
advanced ground-based telescopes and instruments.  The average of all data from both 
nights yields S/N ratios in the range $10-17$ (average of 14) at $H$ and $10-26$ (average 
of 18) at $K$, estimated using 12 sections of 50 pixels each over the spectral ranges plotted 
in Figure \ref{fig2}.  Without any inter-order adjustment, the relative levels of the three shortest 
wavelength orders appear to match each other well in addition to being consistent with the $J$ 
and $H$ photometry.

In order to perform a check on the flux level of the $K$-band spectroscopy, the expected 
photometric flux in the $K$-band was calculated based on the timing of the spectroscopic 
observations over the known $K$-band variability represented in Figure \ref{fig3}.  Taking 
$K_{min}=17.37$ mag from Table \ref{tbl2} and adding to that the $-0.70$ mag difference 
between the average $K$-band minimum and maximum from \citet*{deb06}, yields $K_
{max}=16.67$ mag.  From Figure \ref{fig3}, the spectroscopic observations span, over time, 
an estimated average minimum to maximum ratio of 5:7, yielding a flux-averaged magnitude
of $K_{avg}=17.02$ mag over all the exposures.  All three of these $K$-band fluxes are plotted 
in Figure \ref{fig2}.  As can be seen in the figure, the spectroscopic flux over the $K$-band is 
commensurate with expectations without any adjustments.

The spectrum in Figure \ref{fig2} appears mostly consistent with a Rayleigh-Jeans slope 
in the $J$-band, where an emission line is likely detected at Paschen $\beta$ and possibly 
also at Paschen $\gamma$.  It is unclear whether any emission originates on the companion 
via heating by the primary, or on the white dwarf via accretion; both are possible.  A modest 
excess is detected over the $H$-band, and is consistent with the Table \ref{tbl2} photometry.
At $K$-band there is a definite excess throughout the window and four possible noteworthy 
features with varying degrees of uncertainty: a slight peak near 2.1 $\mu$m that could be a 
cyclotron harmonic, Brackett $\gamma$ emission, the CO bandhead at 2.29 $\mu$m, as well 
as a significant upturn in slope near 2.3 $\mu$m.

\section{ANALYSIS}

\subsection{Binary Components}

The near-infrared spectrum of SDSS 1212 has been fitted with several spectral models and 
templates.  The binary components were analyzed following the method of \citet*{dob05}.  A 
model white dwarf spectrum spanning $0.3-10$ $\mu$m was generated for a non-magnetic, 
pure hydrogen atmosphere at $T_{\rm eff}=10,000$ K and log $g=8.0$ \citep*{sch05a}, using 
{\sf tlusty}\footnote{http://nova.astro.umd.edu} \citep*{hub95} and {\sf synspec}$^1$.  Since the 
white dwarf has a significant magnetic field, with Zeeman-split Balmer lines in its optical spectrum, 
it is not possible at present to more precisely determine its parameters by the usual method of 
Balmer line modeling.  The model spectrum has been normalized to the $g$ magnitude of the 
white dwarf and smoothed to match the spectral resolution of the GNIRS data.

Template $1-2.3$ $\mu$m spectra for brown dwarfs of late L and early T type were taken from 
low resolution Keck / NIRSPEC datasets \citep*{mcl03}.  These templates were then extended 
to 2.5 $\mu$m using similar near-infrared UKIRT / CGS4 spectral data \citep*{geb02,leg01}, in 
order that they span the full extent of the GNIRS wavelength coverage.  The NIRSPEC spectral 
templates have similar resolution ($R\approx2000$) to the GNIRS data and hence no smoothing 
of either data was performed.  However, the CGS4 data had to be interpolated between resolution 
elements in order to create sufficient data points to compare with the GNIRS spectrum.  Empirical 
model fluxes, determined from near-infrared photometry and parallaxes \citep*{tin03} were scaled 
appropriately to $d\approx150$ pc, the nominal distance to a white dwarf with the assumed 
parameters listed above.

\subsection{Cyclotron Emission}

A physically realistic model for cyclotron emission was calculated utilizing the formulation and 
methods described in detail by \citet*{hoa07b} and \citet*{bri07}.  The model employs five critical 
parameters to produce the emergent cyclotron spectrum.  The magnetic field strength and electron
temperature determine the cyclotron harmonic wavelengths, while the viewing angle affects the 
overall shape of the peaks.  Also included are a parameter related to the transition from optically 
thick to optically thin emission (typically well-constrained by the observations), and a scaling factor
linked to the area of the emitting region; a few percent of the white dwarf surface area for the models 
shown here.

Initial attempts to fit the spectrum of SDSS 1212 using cyclotron emission in a $B=13$ 
MG \citep*{sch03} field were problematic.  The sum of the white dwarf and secondary model 
components does a decent job of reproducing the phase-averaged spectrum shortward of 
$K$-band.  Yet the addition of such a cyclotron model component in this bandpass does not 
account for the upturn near 2.3 $\mu$m in the GNIRS data.  Several circumbinary dust models 
were considered to explain this observed rise in flux at the end of the $K$-band, but were 
subsequently discarded upon the realization that this feature must be a cyclotron harmonic 
for the following reasons.

As pointed out by \citet*{deb06}, the zero-temperature $m=4$ cyclotron harmonic for a 13 MG 
magnetic field lies near 2.1 $\mu$m.  But in such a model, this cyclotron feature alone, varied 
by 100\%, is insufficient to produce the peak-to-peak flux change ($\Delta m\approx1$ mag) 
seen in their $K$-band light curve.  Once the contributions of the white dwarf and secondary are
accounted for, {\em all} the remaining flux in the $K$-band must be 100\% variable in order to 
explain the observed light curve; leaving cyclotron as the only viable candidate for this extra 
emission.  Furthermore, a 13 MG magnetic field should produce cyclotron harmonics and 
photometric variability within the $H$-band ($m=5$ would be near 1.6 $\mu$m), as seen in 
EF Eridani and AM Herculis \citep*{kaf05,har04}.  The resolution to this dilemma is that the 
magnetic field of SDSS 1212 must be less than 13 MG, such that more cyclotron emission is 
present within the $K$-band, while the harmonics at shorter wavelengths are diminished.

A magnetic field of 7 MG should have two zero-temperature cyclotron harmonics within the 
$K$-band, near 2.2 and 2.5 $\mu$m for $m=6$ and 7 respectively.  Although \citet*{sch03} 
determine an equivalent dipolar field of $B_d=13$ MG for SDSS 1212 (via models and assumed 
geometry), the mean surface field as determined by the measured Zeeman splitting of H$\alpha$ 
and H$\beta$ is $B_s=7$ MG \citep*{sch05a}.  As shown below, this magnetic field strength is 
consistent with the absence of observable cyclotron harmonics in $H$-band, and the presence 
of two harmonics in the $K$-band which together can explain the large photometric variability 
seen in this filter.

Figures \ref{fig4} and \ref{fig5} display the most successful model reproductions of the GNIRS 
cross-dispersed spectrum of SDSS 1212.  The essential difference between the two figures is 
that an L8 dwarf template was used for the Figure \ref{fig4} plots, while a T2 dwarf template was
used in the plots of Figure \ref{fig5}.  It should be noted that between these two figures, a minor
scaling readjusment was made to the binary components in order to best reproduce the observed 
spectrum of SDSS 1212.  This rescaling amounted to 8\%, corresponding to a 4\% change in the 
assumed distance to the system; well within the 14\% minimum uncertainty arising from its poorly 
constrained temperature and unknown surface gravity \citep*{bur06a,sch05a}.

The upper panels in both figures represent the best fit to the spectrum with only the binary 
components at $d\approx150$ pc; the white dwarf model plus a brown dwarf template.  Already 
it is clear that the $H$-band excess is largely reproduced by the addition of the L8 dwarf, but not 
as well by the T2 dwarf.  However, in both upper panels there appears to be excess flux over the 
$1.05-1.20$ $\mu$m range compared to the model.  It is not clear if this is the result of poor signal 
or calibration in order 6 of the cross-dispersed GNIRS data, or a real detection.  

The middle panels of both figures add a single field cyclotron emission component to the 
upper panel model fits.  In both figures, the models are for a 7.6 MG magnetic field, a 5 keV 
plasma temperature, and a viewing angle of $75\arcdeg$, with the remaining parameters varied 
slightly so that the peaks of the cyclotron harmonics match the phase-averaged $K$-band flux.  
Although this set of models can account for a significant portion of the $K$-band flux and hence
go a long way towards accounting for the photometric variability in this window, they fail miserably 
in the region between the cyclotron peaks.  It is perhaps noteworthy that the T2 model requires a 
stronger cyclotron component than does the L8 model (due to the bluer $H-K$ color of T dwarfs 
versus L dwarfs) and is more readily consistent with the $K$-band variability.

The lower panels in both figures display cyclotron models similar to the middle panels, 
but with a viewing angle of $50\arcdeg$; a value commensurate with the determination 
by \citet*{bur06a}.  Although this angle is not strongly constrained by their data, the GNIRS 
spectra appear to demand an intermediate viewing angle, whose result is to broaden the 
cyclotron harmonics, effectively blurring them together for low viewing angles \citep*{hoa07b}.  
In Figure \ref{fig4}, the magnetic field employed is 7.0 MG , while in Figure \ref{fig5} it is 7.2 
MG, both adjusted slightly to best fit the overall spectrum when combined with their respective 
secondary models.  Both these model fits do a decent job of reproducing the entire spectrum 
from $1.2-2.5$ $\mu$m, but the L8 model matches the $H$- and $K$-band data more closely 
than the T2 model, as in the middle panels.

\subsection{Substellar Secondary}

The three component, composite model fits contain a sufficient number of free parameters 
that their success in reproducing the $1.2-2.5$ $\mu$m spectrum of SDSS 1212 may not 
seem surprising, but is nonetheless informative.  During initial attempts to model the GNIRS
spectrum, secondary spectral templates of types L6 and T5 were shown to be significantly
too bright and too dim, respectively, beyond 1.5 $\mu$m when their overall $J$-band flux 
level was matched to the data.  Figure \ref{fig4} is consistent with the direct spectroscopic 
detection of a substellar object with spectral type L8.  While the T2 model in Figure \ref{fig5} 
appears somewhat inconsistent with the data, it does not rule out an earlier type T dwarf, for 
which available templates were lacking.  Hence, the companion to SDSS 1212 has a likely 
spectral type between L8 and T1.  Uncertainty in the distance to the white dwarf allows for 
some leeway in the relative contribution of the binary components at near-infrared wavelengths.  
While its poorly constrained effective temperature permits photometric distances both nearer or 
farther than $d\approx150$ pc, the unconstrained mass and radius of the white dwarf are likely 
to overpredict its distance.  A typical isolated white dwarf surface gravity, or log $g=8.0$ (0.60 
$M_{\odot}$ at $T_{\rm eff}=10,000$ K; \citealt*{ber95}), is more likely to be too low than too 
high, especially if the primary has been accreting material over Gyr timescales.  This would 
tend to reduce the photometric distance and lead to later spectral types for the companion.  If 
SDSS 1212 is more massive than log $g=8.0$, there is a greater likelihood that the companion 
has an early T spectral type.  The lower panel in the L8 model, which also matches the observed 
$K$-band spectrum well, seems to indicate that the CO bandhead at 2.29 $\mu$m may have 
been detected, albeit just barely.  If correct, it would be further evidence that the companion is a 
late L dwarf, since this CO feature weakens significantly in T dwarfs \citep*{cus05,mcl03,geb02}.  

\subsection{The Nature of SDSS 1212}

The two models for the nature of mass transfer and accretion in SDSS 1212 are Roche lobe 
overflow and wind capture.  The near-infrared observations do not distinguish between these 
two evolutionary scenarios, but are informative for the following reason.  The confirmed presence 
of cyclotron radiation adds one more observational property that SDSS 1212 now shares with 
polars (especially those with likely substellar secondaries in persistent low states such as EF 
Eridani).  The traits it shares are: an appreciable x-ray luminosity, the presence of an accretion 
hot spot, a variable optical light curve with a similar shape to those of known polars in low states 
(i.e. with larger modulations in the blue and ultraviolet compared to the red), a spin-locked binary
configuration, and cyclotron emission \citep*{bur06a,sch05a}.  Occam's razor would argue that 
SDSS 1212 is a polar.

Furthermore, the hot spot x-ray luminosity of SDSS 1212 is $L_x\approx3\times10^{29}$ 
ergs s$^{-1}$, which is nearly identical to that measured for EF Eridani during its 10 yr low;
namely $L_x\approx2\times10^{29}$ ergs s$^{-1}$ \citep*{sch07}.  While the x-ray emission 
from many suspected, detached, wind capture systems may be too weak (and more importantly, 
not periodically variable) to originate via accretion onto the white dwarf, but rather from activity 
on the M dwarf secondary, L dwarfs simply do not have sufficient x-ray luminosities, hence 
favoring the low state polar model for SDSS 1212  \citep*{bur06a}.  Also, there is the issue of 
how such a low mass object generates sufficient wind to be captured by the magnetic white 
dwarf.  To date, no model has emerged which can explain how an old L or T dwarf might 
generate a wind on the order of $10^{-13}M_{\odot}$ yr$^{-1}$.

\section{DISCUSSION}

\subsection{Past and Future Evolution}

The gross evolution cataclysmic variables should consist of three fundamental phases:

(1) A short- to long-lived phase as a detached main sequence binary which subsequently 
evolves to shorter orbital periods through common envelope evolution as the primary leaves 
the main sequence and becomes a first ascent or asymptotic giant.  Angular momentum is 
transferred, via friction, into the envelope itself, ejecting it in the process \citep*{pac76}.  If 
the orbit is still too large for the secondary to make contact with its Roche lobe, then three 
mechanisms can bring this about, in principle: gravitational radiation, magnetic braking, or 
nuclear evolution.  The timescales of these processes depend on the masses involved, but 
if one assumes a canonical 0.6 $M_{\odot}$ white dwarf and a 0.2 $M_{\odot}$ secondary 
(corresponding to the mass distribution peak among detached, unevolved, low mass 
companions to white dwarfs; \citealt*{far05a}), then both gravitational radiation and magnetic 
braking operate within Gyr timescales for period ranges $P\la3$ hr and $P>3$ hr respectively, 
whereas nuclear evolution will play no role within this timeframe \citep*{pat84}.  

(2) A phase in which the shrinking Roche lobe of the secondary becomes smaller than its 
radius, causing unstable mass transfer onto the white dwarf.  In principle this is a runaway 
inspiral until the mass ratio is close to one, but if the secondary has $M<0.5$ $M_{\odot}$ to 
begin with (as it should in the vast majority of cases based on the aforementioned companion 
mass distribution), then conservation of angular momentum competes to partially counterbalance 
the losses due to any braking (magnetic or tidal) and gravitational radiation \citep*{nel01}.  At 
some point the trend reverses outward as the orbital evolution becomes dominated by 
conservation of angular momentum and any mass lost from the system entirely (e.g. 
nova-like outbursts, accretion-driven winds).

(3) A final, debatable, phase consisting of a once more detached or very low accretion 
rate semi-detached binary comprised of two degenerate objects in an expanded orbit 
which persists (now subject only to gravitational radiation) up to 10 Gyr or so, depending 
on the final masses and separation.  This eventual state depends on the thermal and nuclear 
history of the secondary but requires a degenerate secondary core in order to cease expansion 
permanently.  Thus, a system may end its existence as a white dwarf with a close, altered 
substellar companion, both objects cooling gradually toward oblivion.

The ultimate fate of catacylsmic variables outlined above is uncertain for two fundamentally 
different reasons.  The first issue is that the outcome itself is unknown, with end predictions 
for secondaries including; complete stripping, tidal disruption, persistence as semi-detached 
substellar donors with very low mass transfer rates, and survival as detached substellar or even 
planetary mass objects \citep*{lit03,kol99,pat98,how97}.  The second issue is detectability, which 
itself divides into two categories: searching for either erupting systems or their burnt out remnants.

\subsection{Recently Deceased or Born Dead?}

So far, astronomers who study cataclysmic variables have concentrated primarily on 
those systems known to be eruptive in order to search for and understand possible end states, 
including substellar companions.  The means of identifying such low mass secondary systems 
relies heavily on semi-empirical and theoretical relations between observed periods and spectral 
type, mass ratio, effective temperature, and mass transfer rate \citep*{lit06,lit07,har05,pat05,kol99,
how97}. Without models, there is no means of determining whether a given substellar companion 
is migrating inward or outward, i.e. approaching or leaving its period minimum.  With the prior
history of these interacting systems unknown, the beginning and end states may appear 
identical, theoretically.

Additionally, it is difficult to know with certainty what the size of the seconday star is in such 
systems, especially if the mass transfer rate is low -- consistent with quiescence or an underfilled 
Roche lobe (i.e. a non-catacylsmic variable).  If a companion does indeed have a radius smaller 
than its Roche lobe, then strictly speaking it is detached, and the question remains: is Roche lobe 
contact in its future or in its past?  The recently identified low mass transfer magnetic binaries 
reported in \citet*{sch05b} are good candidates for being detached due to the combination of 
the inferred accretion rates, orbital periods, and extremely low or absent x-ray emission.  {\em But 
if EF Eridani can turn off for nearly 10 yr, then so can other polars} \citep*{sch07}.  SDSS 1212
sits somewhere on the border between a typical polar and the low accretion rate systems, having 
respectable x-ray accretion luminosity and a relatively low inferred mass transfer rate \citep*{bur06a}.  
If one assumes the substellar companion to SDSS 1212 has a mass and effective temperature 
similar to WD 0137$-$349B (L8V, see Table \ref{tbl3}; \citealt*{bur06b}), then its unperturbed radius 
should be 0.09 $R_{\odot}$ for ages between $1-5$ Gyr \citep*{bar03}, but its Roche lobe should 
be located roughly at 0.11 $R_{\odot}$ for a 0.60 $M_{\odot}$ primary.  Therefore, the companion 
is either: 1) recently deceased, far out of thermal equilibrium and inflated as expected for an object 
past period minimum with the implied long history of mass-radius imbalance; or 2) born dead, 
intrinsically detached, and slamming the magnetic white dwarf with all of its (irradiatively driven?) 
wind.  This raises similar questions about the nature of all other low accretion rate magnetic 
binaries.

Depending on the past thermal and nuclear history of the secondary, especially if the core is 
degenerate, and how much mass escapes the binary completely and adiabatically, it is quite 
possible for the orbital period to increase significantly over the minimum into a region where 
no interaction (or wind accretion only) occurs \citep*{how97,jea24}.  Without any distinct spectral 
features or mass-radius anomalies in such detached binaries, there may be little or no means to 
differentiate empirically between the pre- and post- states of these innately interacting systems. 

\subsection{The Frequency of the Doubly Dead}

Objects must first be detected as a catacylsmic variable in order to look for substellar 
companions; hence, selection effects play a large role in the number of systems suspected to 
harbor such cool degenerate companions \citep*{lit03}.  However, this overlooks the obvious 
question: if the end states are detached double degenerate systems, where are they?  If one 
restricts a search to eruptive systems, then the potential for success is limited from the beginning 
for reasons stated above.  Why not search directly for systems which are detached and which may 
represent final states?  This seems to be the only way to reconcile the well-known discrepany 
between the number of predicted objects past period bounce and the number observed or 
suspected from observation \citep*{pat98,how97}.

Clearly, the best place to look for these types of end states is in the infrared where cool, 
Jupiter-sized companions easily outshine their tiny white dwarf primaries \citep*{far05a,far05b}.  
There are only two issues associated with a search around white dwarfs for extinct cataclysmics 
-- the white dwarf must first have been identified or previously known and one must have sufficient 
sensitivity to detect its cooling leftovers.  If the end state white dwarfs are very cool and far away 
such that $V>21$ mag (e.g. $T_{\rm eff}<5000$ K and $d>100$ pc for $M=0.9$ $M_{\odot}$), 
then only the Sloan Digital Sky Survey will detect them but will be quite insensitive to any 
substellar companions.  While the above conditions may not be representative, given that 
cataclysmic variables appear to be roughly 100 times less abundant than white dwarfs 
\citep*{pat84}, a large sample of targets would be required to perform a statistically rigorous 
search; if all cataclysmic variables are in this state, a search would only yield 1 per 100 white 
dwarfs (not very encouraging).

If these close but detached systems exist in any significant numbers, they have not yet been 
identified from various surveys of relatively nearby white dwarfs.  Spatially unresolved L type 
companions, corresponding to $0.05-0.07$ $M_{\odot}$ for unevolved ages of $1-5$ Gyr, are 
rare, with only 3 known among well over $N=1000$ targets \citep*{hoa07a,bur06b,far05a,far04,
far04b,wac03}.  Spatially unresolved T type companions, corresponding to $0.02-0.05$ 
$M_{\odot}$ for unevolved ages of $1-5$ Gyr, are also rare, with none detected or suspected 
in various surveys totalling $N\approx200$ targets (\citealt*{mul07,han06,far05a}; Farihi, Becklin, 
\& Zuckerman 2008, in preparation).  It is possible that stellar evolution has yet to produce such 
systems; i.e. the initial mass function for secondaries in such systems, convolved with the finite 
age of the Galaxy, implies that the vast majority of these systems will continue in phases (1) and 
(2) for up to 10 Gyr.  It could be that the lack of detected end states is an observational bias; if 
these systems are still marginally semi-detached then the signature of the secondary star could 
be masked by the accretion disk in dwarf novae and by cyclotron emission in polars. Alternatively, 
phase (3) never occurs and the companion is cannibalized, evaporated, stripped or tidally 
destroyed.  This would also explain the lack of observed end states.

For completeness, something ought to be said about the empirical frequency of zero age 
cataclysmic variables with a brown dwarf secondary \citep*{pol04}.  Since their pre- states 
would consist of a white dwarf with a detached substellar companion, the above arguments
and evidence apply ($f<0.5$\%).  Additionally, the progenitors of such pre- states would be 
main sequence stars with brown dwarf companions in the few AU range, which are also 
extremely rare, occuring with a well-constrained frequency, $f<0.5$\% \citep*{but06,mar05}.  
Hence, there should be no doubt that cataclysmic variables form with brown dwarf 
secondaries quite rarely.

\subsection{Detached Companions to Magnetic White Dwarfs}

The same argument which applies to detached substellar companions to white dwarfs 
also applies to detached stellar companions to magnetic white dwarfs, although in this latter 
case finding them should be easier in principle \citep*{lie05}.  One might suggest a sensitive 
search for magnetism among known lists of close but detached, and wide common proper 
motion companions.  Large lists of these white dwarf plus red dwarf systems exist in the following 
references: for known radial velocity pairs see \citet*{mor05}, for spatially unresolved pairs see 
\citet*{hoa07a,far06,far05a,wac03}, and for wide common proper motion pairs see \citet*{far05a,
sil05}.  All of these lists contain targets which are likely to be superior to white dwarfs from the 
Sloan Digital Sky survey since they are brighter and, hence, more sensitive searches could be 
conducted. However, it should be noted that the main observational indicators of a magnetic 
white dwarf, Zeeman splitting and circular spectropolarimetry, are strongest in H$\alpha$ and 
H$\beta$.  Unfortunately, in unresolved white dwarf plus red dwarf binaries (i.e., the potential
cataclysmic variables) these features, especially H$\alpha$, are often completely filled-in by 
the red dwarf.  This may be another reason for the curious lack of detected pre-polar systems 
to be considered along with those already discussed by \citet*{lie05}.

\section{FUTURE WORK}

Unfortunately, the intrinsic faintness of SDSS 1212 precludes several interesting avenues 
of follow up, such as phase-resolved near-infrared spectroscopy or mid-infrared spectroscopy.  
Ultraviolet and x-ray observations are currently planned to firmly distinguish between standard 
mass transfer through Roche lobe overflow versus accreted wind.  High resolution optical 
spectra of SDSS 1212 may reveal the presence of elements accreted from its companion.  
Such observations could, in principle, distinguish between Roche lobe overflow mass transfer 
and wind accretion as well.

There may be little or no hope of directly detecting the bare substellar companion beyond 
the $H$-band spectrum presented here.  The 7 MG cyclotron models used successfully in this
paper predict the fundamental harmonic wavelength should be 15.3 $\mu$m, thus potentially
swamping every instrument aboard {\em Spitzer} which might otherwise be capable of directly 
detecting the flux of an L8 dwarf at $d=150$ pc; namely IRAC $3-8$ $\mu$m photometry, and 
IRS 16 $\mu$m peak-up imaging.  However, IRAC observations should directly test the lower 
magnetic field cyclotron models employed to fit the GNIRS data, which predict relatively strong 
emission near 3.1, 3.8, 5.1, and 7.7 $\mu$m, corresponding to the zero-temperature $m=5$, 4, 
3, and 2 harmonic numbers in a 7.0 MG magnetic field.  Yet the possible presence of circumbinary 
dust emitting at IRAC wavelengths, as seen in several polars \citep*{hoa07b,bri07,how06}, may 
complicate the interpretation of any mid-infrared photometry for SDSS 1212.

\acknowledgments 

This paper benefitted greatly from the insight of an anonymous referee, to whom the authors 
are commensurately grateful.  The authors would like to thank the following people: B. Rodgers, 
G. Doppman, and K. Labrie (for much appreciated help in understanding and processing GNIRS 
data); A. Stephens (for NIRI particulars); P. Hirst (for WFCAM particulars); S. Leggett (for a useful 
discussion regarding photometric calibration errors).  M. R. Burleigh is supported by a UK Science 
and Technology Facilities Council Advanced Fellowship.  Spectroscopic observations for this paper 
were taken as part of the Gemini Director's Discretionary Time GS-2006A-DD-5.  Photometric 
observations were obtained as part of Gemini Queue Program GN-2006A-Q-51 and UKIRT Service 
Program 1653.  Gemini Observatory is operated by the Association of Universities for Research in 
Astronomy, Inc., under a cooperative agreement with the NSF on behalf of the Gemini partnership: 
the National Science Foundation (United States), the Particle Physics and Astronomy Research 
Council (United Kingdom), the National Research Council (Canada), CONICYT (Chile), the 
Australian Research Council (Australia), CNPq (Brazil) and CONICET (Argentina).  The United 
Kingdom Infrared Telescope is operated by the Joint Astronomy Centre on behalf of the UK Particle 
Physics and Astronomy Research Council.  This publication makes use of data products from the 
Two Micron All Sky Survey, which is a joint project of the University of  Massachusetts and the Infrared 
Processing and Analysis Center / California Institute of Technology, funded by the National Aeronautics 
and Space Administration and the National Science Foundation.  This paper has made use of the 
Aladin Sky Atlas operated at CDS, Strasbourg, France.

{\em Facility:} \facility{Gemini (GNIRS, NIRI)}, \facility{UKIRT (WFCAM)}

\clearpage

\begin{deluxetable}{ccccccc}
\rotate
\tablecaption{WFCAM Photometric Calibration Data\label{tbl1}}
\tablewidth{0pt}
\tablehead{
\colhead{Number}      					&
\colhead{2MASS Name}  					&
\colhead{$H$ (mag)}   					&
\colhead{$K_s$ (mag)} 					&
\colhead{$H_0$\tablenotemark{\dag} (mag)} 	&
\colhead{$K_0$\tablenotemark{\dag} (mag)} 	&
\colhead{Remarks}}

\startdata
1	&J12114360+0137020	&$13.841\pm0.035$	&$13.701\pm0.050$		&23.032		&22.354		&\\
2	&J12114583+0139374	&$13.435\pm0.030$	&$13.337\pm0.038$		&23.083		&22.391		&\\
3	&J12114736+0133594	&$14.098\pm0.039$	&$14.105\pm0.063$		&23.031		&22.426		&\\
4	&J12114758+0137223	&$14.021\pm0.037$	&$14.062\pm0.062$		&23.040		&22.479		&\\
5	&J12115403+0132049	&$13.991\pm0.046$	&$13.766\pm0.058$		&23.069		&22.325		&\\
6	&J12115541+0130399	&$13.431\pm0.027$	&$13.299\pm0.035$		&23.041		&22.356		&\\
7	&J12120408+0135365	&$15.046\pm0.066$	&$14.776\pm0.123$		&23.076		&22.282		&1,2\\
8	&J12120582+0135157	&$14.987\pm0.072$	&$14.955\pm0.138$		&22.980		&22.404		&1,2\\
9	&J12120662+0138153	&$14.065\pm0.053$	&$13.974\pm0.068$		&23.075		&22.405		&\\
10	&J12120989+0135259	&$15.894\pm0.174$	&$15.538\pm0.232$		&22.908		&22.115		&1,2\\
11	&J12121658+0137042	&$14.290\pm0.041$	&$14.137\pm0.066$		&23.039		&22.452		&2\\
12	&J12121667+0135257	&$13.868\pm0.022$	&$13.805\pm0.068$		&23.036		&22.383		&2\\
13	&J12122238+0130101	&$12.927\pm0.022$	&$12.904\pm0.032$		&23.058		&22.410		&\\
14	&J12122266+0137425	&$13.602\pm0.041$	&$13.536\pm0.055$		&23.011		&22.470		&\\
15	&J12122647+0140477	&$13.040\pm0.029$	&$13.030\pm0.037$		&23.070		&22.412		&\\
\enddata

\tablecomments{
(1) Not used in the present paper;
(2) Used in \citet*{deb06}}

\tablenotetext{\dag}{Transformations between 2MASS and UKIRT photometric systems 
were ignored (see \S3.1).}

\end{deluxetable}

\clearpage

\begin{deluxetable}{cccccc}
\tablecaption{Photometry of SDSS 1212\label{tbl2}}
\tablewidth{0pt}
\tablehead{
\colhead{Passband}      			&
\colhead{Instrument}      			&
\colhead{Magnitude (mag)}      		&
\colhead{$\sigma_{phot}$ (mag)}    	&
\colhead{$\sigma_{cal}$ (mag)}      	&
\colhead{$\sigma_{total}$ (mag)}}

\startdata

$H$	&WFCAM	&17.676		&0.089	&0.021	&0.091\\
$H$	&NIRI	&17.710		&0.064	&0.030	&0.071\\
$K$	&WFCAM	&17.372		&0.105	&0.047	&0.114\\
$K$	&NIRI	&16.693		&0.033	&0.030	&0.045\\

\enddata

\tablecomments{An intrinsic calibration uncertainty of 3\% was assumed for the NIRI data.  
See Figure \ref{fig1} for the phases corresponding to the photometric observations.}

\end{deluxetable}

\clearpage

\begin{deluxetable}{lcccccc}
\tablecaption{Close Binary Parameters for SDSS 1212 and WD 0137$-$349\label{tbl3}}
\tablewidth{0pt}
\tablehead{
\colhead{Star}      				&
\colhead{$P$ (min)}    			&
\colhead{$M_1$ ($M_{\odot}$)}   	&
\colhead{$M_2$ ($M_{\odot}$)}   	&
\colhead{$a$ ($R_{\odot}$)}		&
\colhead{$R_2$ ($R_{\odot}$)}   	&
\colhead{$R_{L2}$ ($R_{\odot}$)}}

\startdata

WD 0137$-$349	&115.6	&0.39	&0.05	&0.60	&0.09	&0.13\\
SDSS 1212		&88.4	&0.60	&0.05	&0.57	&0.09	&0.11\\

\enddata

\tablecomments{Values for the white dwarf and secondary masses in the SDSS 1212 
system are assumed since no measurements exist.  However, the resulting semimajor
axis, unperturbed secondary radius, and secondary Roche lobe radius are all relatively 
insensitive to those input values, and reflect likely parameters.  The semimajor axis for 
WD 0137$-$349 was calculated using Kepler's law, yielding a value which differs slightly 
from that reported in \citet*{max06}.}

\end{deluxetable}

\clearpage

\clearpage

\begin{figure}
\plotone{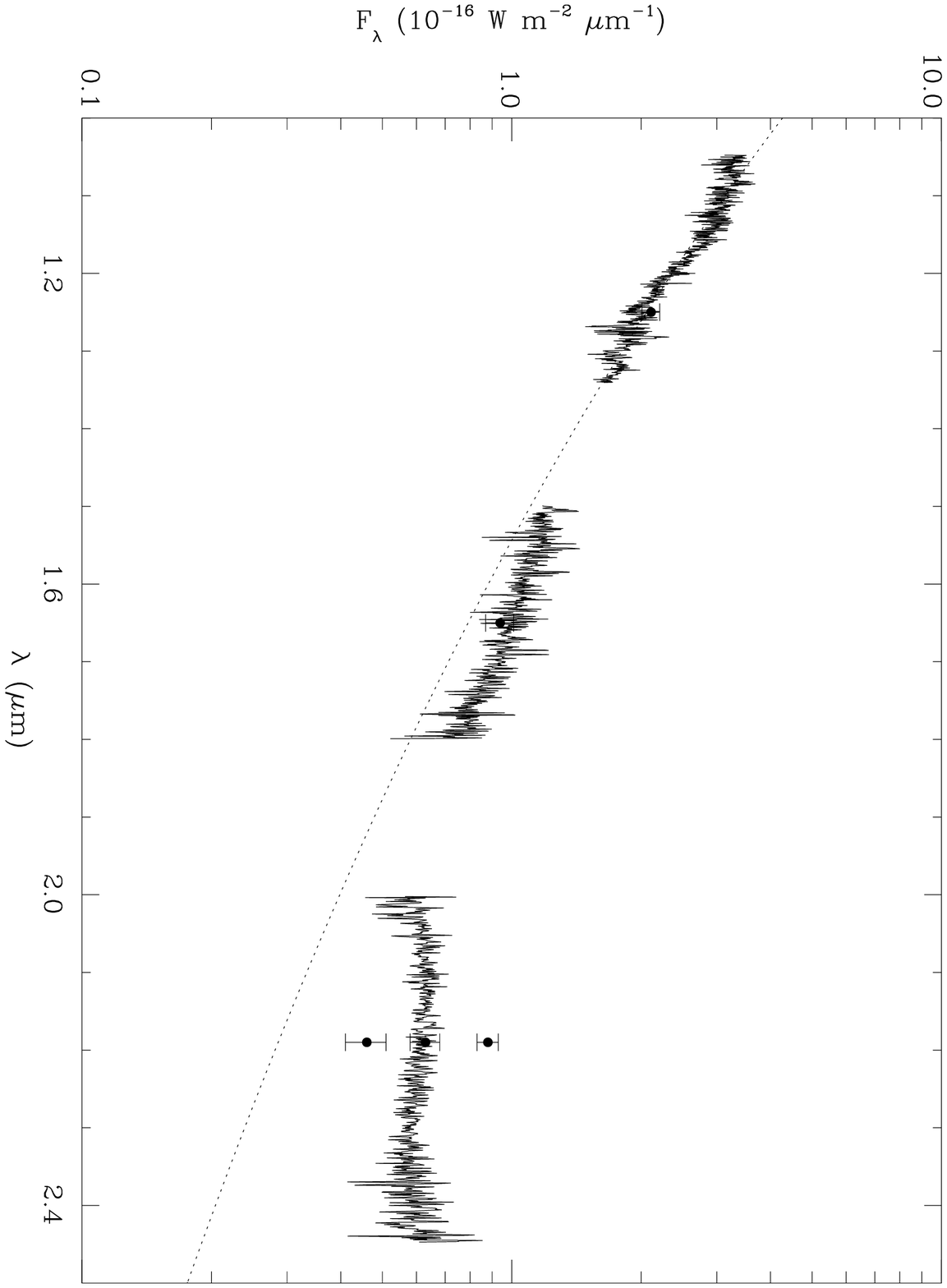}
\end{figure}

\clearpage

\begin{figure}[h]
\caption{Flux calibrated GNIRS spectrum of SDSS 1212, averaged over all nights and 
phases.  Four cross-dispersed orders are displayed within the atmospheric windows.  The 
photometric points and blackbody continuum are identical to that plotted in Figure\ref{fig1}, 
with 2 additional (calculated) $K$-band flux points described in \S3.3.
\label{fig2}}
\end{figure}

\clearpage

\begin{figure}
\plotone{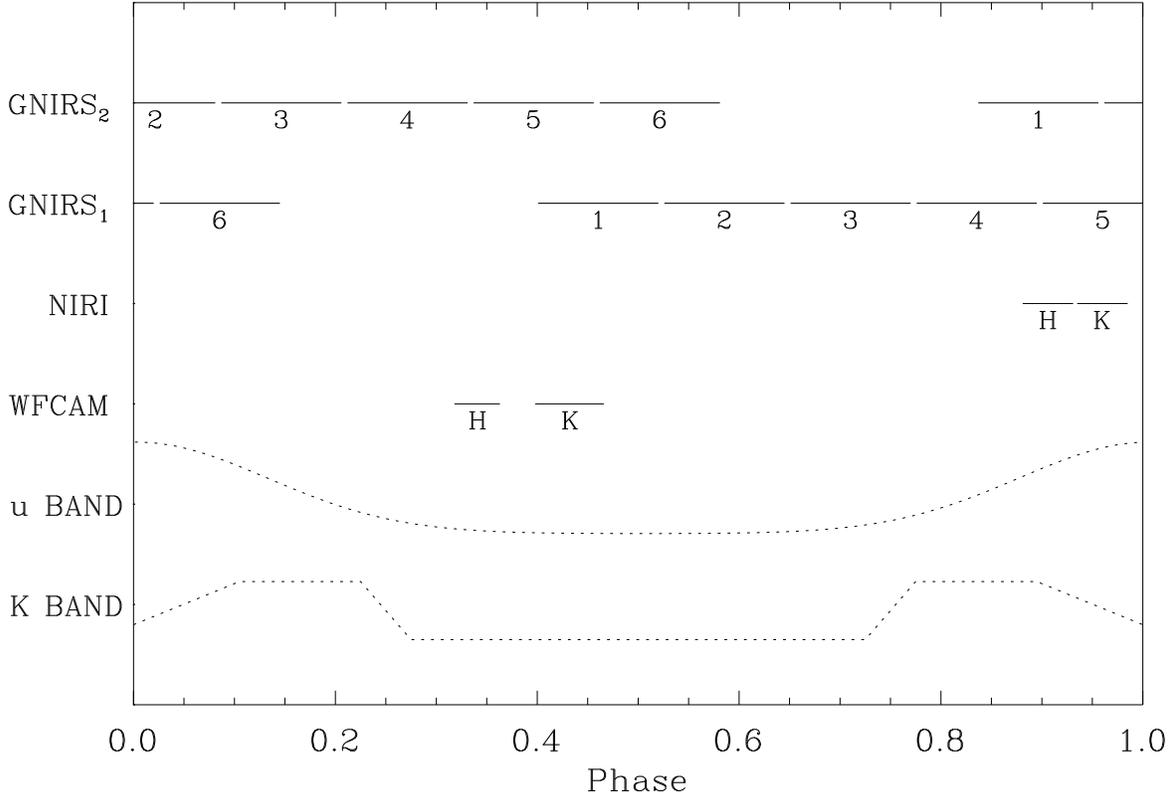}
\caption{Orbital phases corresponding to the photometric and spectroscopic observations 
with each instrument:  GNIRS, NIRI, and WFCAM.  The numbers $1-6$ denote each of 6 
nodded pairs of GNIRS observations taken on two separate nights (subscripted 1 and 2).  
Also shown are the $u$-band light curve model fit of \citet*{bur06a} and the rough $K$-band 
light behavior \citep*{deb06}, both normalized to one.  The relative amplitude of the $u$-band 
peak has been exaggerated by a factor of 10 for visibility.
\label{fig3}}
\end{figure}

\clearpage

\begin{figure}
\plotone{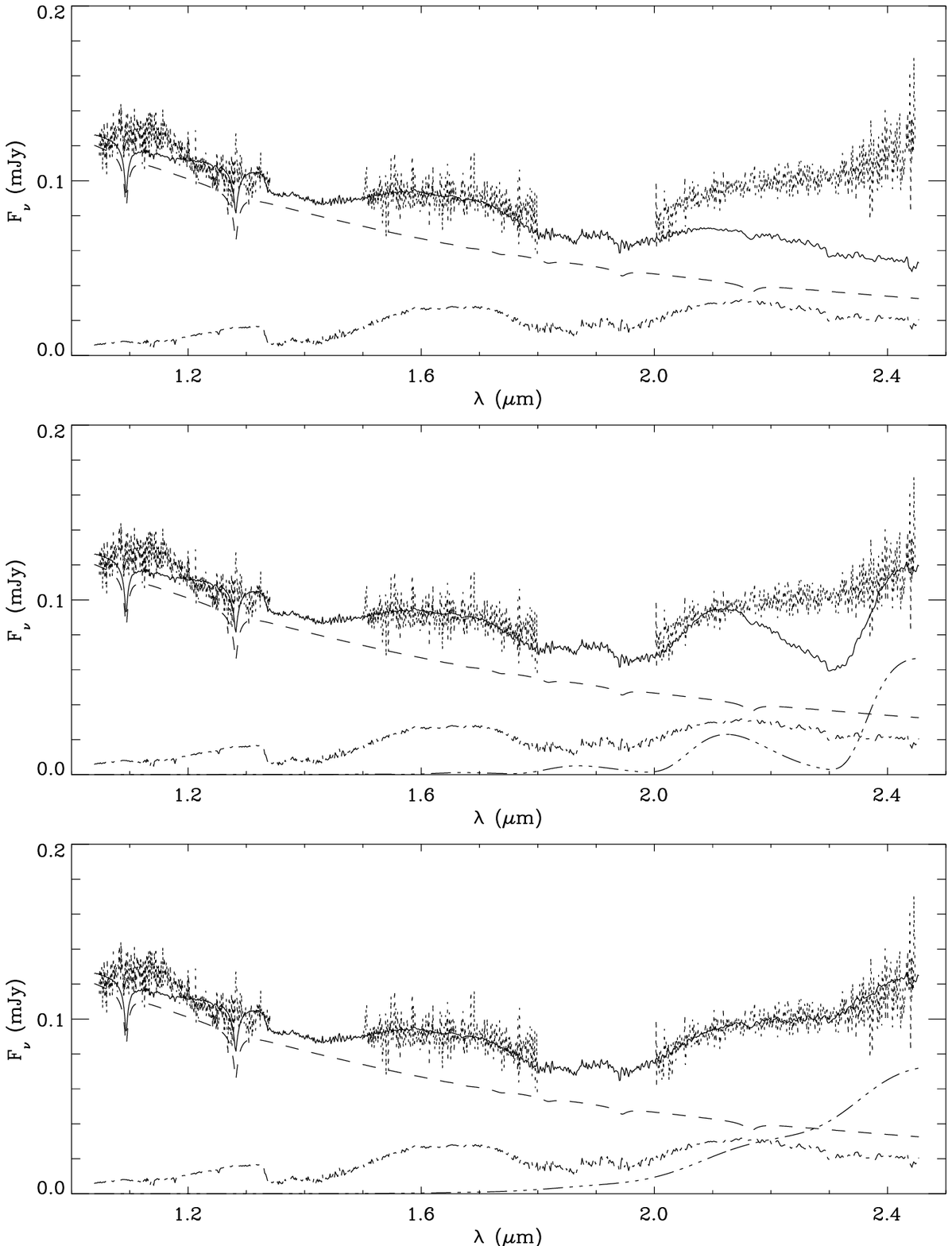}
\end{figure}

\clearpage

\begin{figure}[h]
\caption{Model fits to the GNIRS, phase-averaged, near-infrared spectrum of SDSS 1212.  
In each panel the data are represented by a dotted line and the fit is shown as a solid line.  
Some rescaling has taken place between panels to best fit the data, but the binary components 
remain at $d\approx150$ pc in all plots.  Details of the models and fits are discussed in \S4.1 and 
4.2.  {\em Top panel:}  The dashed line is a pure hydrogen atmosphere, $T_{\rm eff}=10,000$ K 
model for a 0.6 $M_{\odot}$ white dwarf.  The dash dot line is an L8 brown dwarf template 
spectrum.  {\em Middle Panel:}  The dash dot dot line shows the cyclotron model at a viewing 
angle of $75\arcdeg$.  {\em Bottom panel:}  Here, the dash dot dot line shows the cyclotron 
model at a viewing angle of $50\arcdeg$.
\label{fig4}}
\end{figure}

\clearpage

\begin{figure}
\plotone{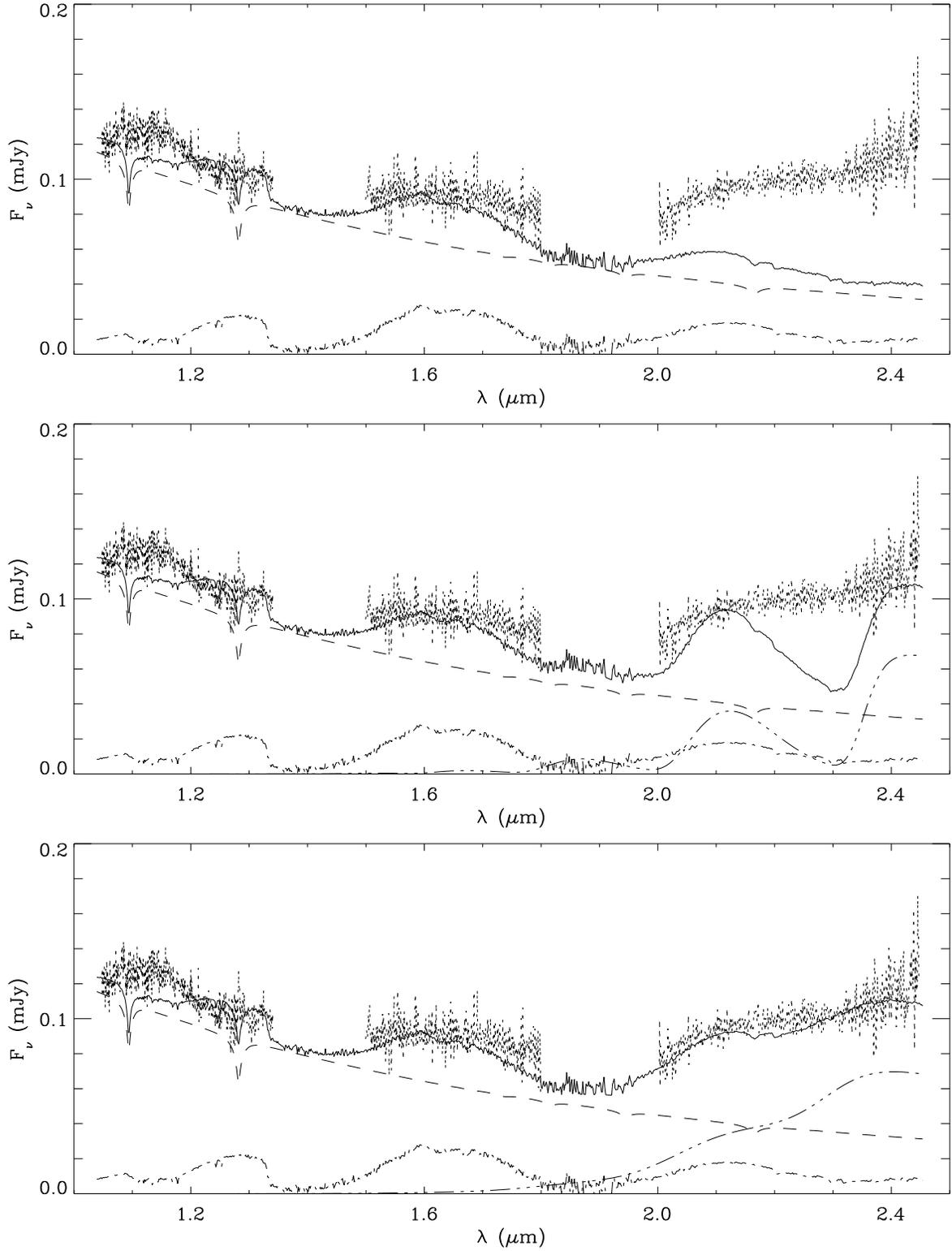}
\caption{Same as Figure \ref{fig4} but for a T2 brown dwarf template spectrum.
\label{fig5}}
\end{figure}

\end{document}